# Thermoelectric quantum oscillations in ZrSiS

Marcin Matusiak[1,*], J.R. Cooper[2], Dariusz Kaczorowski[1]

1. *Institute of Low Temperature and Structure Research, Polish Academy of Sciences,*

   *ul. Okólna 2, 50-422 Wrocław, Poland*

2. *Cavendish Laboratory, Department of Physics,*

   *University of Cambridge, Cambridge CB3 OHE, United Kingdom*

**Topological semimetals are systems in which the conduction and the valence bands cross each other and this crossing is protected by topological constraints. These materials provide an intriguing test of fundamental theory and their exceptional physical properties promise a wide range of possible applications. Here we report a study of the thermoelectric power ($S$) for a single crystal of ZrSiS that is believed to be a topological nodal-line semimetal. We detect multiple quantum oscillations in the magnetic field dependence of $S$ that are still visible at temperature as high as $T$ = 100 K. Two of these oscillation frequencies are shown to arise from 3D and 2D bands, each with linear dispersion and the additional Berry phase expected theoretically.**

*Corresponding author. Tel.: +48-71-3954197; Fax: +48-71-3441029; e-mail: M.Matusiak@int.pan.wroc.pl

Amongst three-dimensional (3D) systems with non-trivial topological states, topological nodal-line (TNL) semimetals, in which Dirac band crossings occur along a line or form a loop in momentum space, attract much attention due to their particularly interesting physics [1]. The representatives of this type of material, examined by angle-resolved photoemission spectroscopy (ARPES), are PtSn$_4$ [2], PbTaSe$_2$ [3], ZrSiS [4-6], ZrSiSe and ZrSiTe [7], and HfSiS [8]. As an outcome of detailed ARPES studies on ZrSiS, this compound has been shown to harbor not only a 3D TNL phase with a diamond-shaped Fermi surface near the Brillouin zone center but also multiple Dirac cones dispersing linearly over an extended energy range up to 2 eV [4-6]. Furthermore, the material hosts linearly dispersive surface states protected by non-symmorphic crystal symmetry [4,5], and is therefore an excellent candidate for the realization of a 2D topological insulator [9,10].

The unique electronic structure of ZrSiS makes it appropriate for comprehensive studies of electrical transport properties, anticipated to reflect the non-trivial character of the bulk states. In accordance with expectations, the magnetoresistance (MR) in ZrSiS has been found extremely large, non-saturating in strong magnetic fields, and highly anisotropic [6,11-13]. However, these features can be interpreted in terms of the semi-metallic nature of this compound with nearly perfect electron-hole compensation and the presence of open orbits at the Fermi surface, i.e. without invoking Dirac states [11]. On the other hand, the MR measurements have also revealed the occurrence of Shubnikov – de Haas (SdH) oscillations, discernible in fields as low as 2 T and at temperatures as high as 20 K [6,11-13]. This finding implies very high mobility of the charge carriers. Fast Fourier transformation (FFT) of the experimental data yielded two characteristic frequencies of about 14 T and 238 T, attributed to a small electron and a larger hole Fermi pocket, respectively, both with the effective mass of the charge carriers close to $0.1m_0$ ($m_0$ stands for the free electron mass) [12]. For the hole pocket, a nontrivial Berry phase π, characteristic of Dirac fermions, has been derived [12], while the electron one has been shown to bear a Berry phase close to zero [6].

The nature of the TNL phase in ZrSiS has recently been probed also via de Haas – van Alphen (dHvA) oscillations [14,15]. These studies confirmed the presence of 3D Dirac states, with a dHvA frequency of about 240 T, high mobility and small effective mass of the hole-type carriers. In addition, a small electron-type Fermi pocket of 2D character has been found that probably corresponds to the Dirac surface states seen by ARPES. Here we demonstrate that measurements of the thermoelectric power is a powerful experimental tool in capturing Dirac states in TNL semimetals. The results obtained for ZrSiS provide novel information on the multiple Dirac bands evidenced in the ARPES experiments on this compound and trace a new path in studying related Dirac materials.

**Results and Discussion**

Despite ZrSiS being dubbed a semi-metal, its thermoelectric power ($S$), as well as the electrical resistivity ($\rho$) [12], measured in absence of magnetic field have a form similar to those for metals. Figure 1 shows that $S$ has a rather moderate value at the room temperature (~ 10 μV/K at 300 K). It stays positive in the entire temperature ($T$) range and varies almost linearly with $T$. The change in slope around $T \approx 50$ K is unlikely to be caused by a phonon-drag contribution, despite the high quality of the crystals studied, because the low temperature part of $S(T)$ shown in the inset in Fig. 1 remains linear, in contrast to the $T^3$ dependence expected from phonon-drag [16]. Perhaps the natural variation in isotope composition for the constituent elements [17] enhances phonon-phonon scattering and suppresses phonon drag. Also, large angle phonon-electron scattering, which is needed to transfer the phonon momentum to the electrons efficiently, will be small at low $T$ when typical phonon wave vectors are large compared to the Fermi wave vectors ($k_F$). The low values of the quantum oscillation frequencies ($F$) show that several carrier pockets are sufficiently small at 10 K. We note in passing that the analogous effect for electron-phonon scattering provides a natural and novel explanation for the $T^3$ dependence of the electrical resistivity observed at low $T$ [12]. Namely the extra factor of $T^2$ arising from large angle electron-phonon scattering that leads to a $T^5$ law in the usual Debye theory, will be absent for small pockets of charge carriers.

While the $S(T)$ dependence of ZrSiS appears to be simple for a material with a multiband electronic structure[4,5] and low charge carrier densities [12], application of a magnetic field along the crystallographic $c$ axis reveals unusual properties. One indication of these is the enormous magnetoresistance [11-13], another is the wide temperature and magnetic field range where quantum oscillations are observed. Fig. 2 shows the field dependence of the thermoelectric power of ZrSiS measured at $T = 1.7$ K. As one can easily judge, the "normal" signal is completely dominated by the oscillatory part that reaches a value of ~8 µV/K at this temperature. Observations of dHvA and SdH oscillations in ZrSiS have been reported recently [13,14], but we find that the thermoelectric response is much more sensitive to these oscillations. This has been found valid for a number of other compounds [18,19]. We are able to detect five different oscillations with frequencies: $F^1 = 8.5$, $F^2 = 15.3$, $F^3 = 57$, $F^4 = 240$ and $F^5 = 583$ T. They can be seen in Fig. 3 which shows the FFT spectrum calculated for data collected at $T = 1.7$ K in the 1 – 12.5 T magnetic field range (here $F^3$ is likely merged with the third harmonic of $F^2$). The $F^3$ and $F^5$ oscillations are reported here for the first time, perhaps because of the exceptional sensitivity of thermoelectric effects to quantization of the Fermi surface mentioned above. The $F^1$, $F^2$, $F^4$ and $F^5$ oscillations are immediately visible in the different field regions of the $S(B^{-1})$ plots shown in the insets of Fig. 2. However, analysis is not straightforward because of the presence of multiple frequencies and their higher harmonics. Figure 4 shows $S(B^{-1})$ plots for selected temperatures, where data in the upper panel has been averaged to remove frequencies higher than ~200 T while the bottom panel show raw (non-filtered) high temperature data. Noticeably, in the high field limit, oscillations with frequency

$F^3 = 57$ T are still present at $T$ as high as 100 K. As the temperature is lowered the oscillations become more pronounced but at the same time less and less regular. It has been suggested that a saw-tooth shape (and therefore high harmonic content) is a typical feature of quantum oscillations in $S$ [18]. It is difficult to definitely judge whether the apparent irregularities in $S(B^{-1})$ dependences at low $T$ are a consequence of a growing contribution from higher harmonics or of a different origin. For instance, the sudden jump in $S$ at $B^{-1} \approx 0.15$ T$^{-1}$ for $T = 1.7$ K marked with an arrow in Fig. 1, and which can also be seen in the filtered data shown in Fig. 4 and Fig. 5, can be an extreme effect of the higher harmonics. However, $B^{-1} \approx 0.15$ T$^{-1}$ also seems to be the field where the highest frequency oscillation ($F^5$) arises and the step in $S(B^{-1})$ could be a sign of a magnetic breakdown. Intriguingly, $B^{-1} \approx 0.15$ T$^{-1}$ is also the field, where the high temperature $F^3$ oscillation (that dominates the high temperature oscillatory signal) emerges. However, magnetic breakdown is expected to give rise to sum or difference frequencies for type I [20] or type II [21] Weyl semimetals respectively. There is no clear evidence for sum or difference frequencies involving either $F^3$ or $F^5$, so therefore, the step-like change in $S(B^{-1})$ probably results from higher harmonics of $F^2$. The clear splitting of the peak at $B^{-1} \approx 0.1$ T$^{-1}$ is probably associated with the Zeeman splitting that was suggested by dHvA studies to be particularly prominent in ZrSiS [14]. Another feature that is unlikely to be related to the emergence of higher harmonics is the sudden drop of $S$ in the high field limit (below 0.093 T$^{-1}$). This anomaly broadens at higher temperatures, but can still be observed even at $T = 66$ K. Here it is worth noting that for the $B > 10$ T, ($B^{-1} < 0.1$ T$^{-1}$) we reach the quantum limit for the $F^1$ band, as implied by the Landau level plot in the inset to Fig. 6. Similar behaviour of the thermoelectric power while crossing from the $n = 1$ to the $n = 0$ Landau level was reported as evidence for massive bulk Dirac fermions in $Pb_{1-x}Sn_xSe$ [22].

As mentioned above, the $F^1$, $F^2$ and $F^4$ oscillations were already seen in previous studies, but intriguingly, while $F^4$ was observed in both dHvA (240 T [14], 237 T [15]) and SdH (238 T

[12], 246 T [6], 243 T [13]) measurements, F[1] was only detected in dHvA (8.4 T [14], 7.8 T [15]), and F[2] only in SdH (14.1 T [12], 18.9 T [6]). Our study shows that while both F[1] and F[2] are present, F[1] generally overlaps with F[2] which is almost a factor of 2 higher. On the other hand, F[1] and F[2] differ in their temperature and field dependences – for example above $B^{-1}$ = 0.5 T$^{-1}$ at $T$ = 1.7 K only F[1] persists, as shown in the main part of Figure 6. Therefore we determine the temperature dependences of the FFT peak amplitudes in three different field regions, 1 – 12.5, 1 – 5, and 5 – 12.5 T, depending on the range where a given frequency is more visible. Figure 5 shows the results of this procedure, where one can see that the evolution of the FFT peak amplitudes clearly differs from the Lifshitz-Kosevich formula that predicts maximal amplitude at zero temperature [23]. This formula cannot be directly applied to oscillations in the thermoelectric power, which, being related to the entropy carried by quasiparticles and their heat capacity, is expected to vanish in the zero temperature limit. Therefore we use the formula for quantum oscillations in $S$ applied recently by Morales et al. to analyse data on UGe$_2$ [24]:

$$A(T) \propto \frac{(\alpha p X)\coth(\alpha p X)-1}{\sinh(\alpha p X)}, \quad (1)$$

As the temperature is lowered, $A(T)$ rises to a maximum at $T \approx 0.11 \, B/(pm^*)$ and approaches zero for $T \to 0$ K. Here $\alpha = 2\pi^2 \, k_B/e\hbar$ ($k_B$ is the Boltzmann constant, $e$ the elementary charge, $\hbar$ the reduced Planck constant), $p$ is the harmonic number, and $X = m^* T/B$ ($m^*$ is the cyclotron mass of charge carriers that is given by the energy dependence of the extremal Fermi surface areas that are in turn proportional to the oscillation frequencies $F$). Fits of the data to Eqn. 1 shown in the lower panels of Fig. 5 give the cyclotron masses summarised in table 1.

**Table 1**

| Frequency (T) | $F^1$: 8.5 | $F^2$: 15.3 | $F^3$: 57 | $F^4$: 240 | $F^5$: 583 |
|---|---|---|---|---|---|
| Effective mass ($m_0$) $S$ (present work) | **0.07** | **0.14** | **0.04** | **0.18** | **1.4** |
| Effective mass ($m_0$) dHvA [14] | 0.025 | - | - | 0.052 | - |
| Effective mass ($m_0$) dHvA [15] | 0.08 | - | - | 0.14 | - |
| Effective mass ($m_0$) SdH [12] | - | 0.1 | - | 0.14 | - |
| Effective mass ($m_0$) SdH [6] | - | 0.12 | - | 0.16 | - |
| Phase shifts $S$ (present work) | 0.04±0.04 | - | 0.38±0.03 | -0.23±0.02 | - |

The only band with an effective mass comparable to $m_0$ is $F^5$, whereas for the others, $m^*$ is one – two orders of magnitude smaller. The values obtained here for $F^1$, $F^2$ and $F^4$ are in good agreement with previously reported results except for estimates given by Hu et al. that are 2 – 3 times smaller [14]. The lightest charge carriers originate from the $F^1$ and $F^3$ bands ($m^* = 0.07$ and 0.04 $m_0$, respectively) so it is especially interesting to see whether the phases of these oscillations exhibit non-trivial properties corresponding to Dirac fermions. In order to do so, we made Landau level fan diagrams, shown as insets in Fig. 5 ($F^1$) and Fig. 6 ($F^2$). Here the integers ($n$) are uniquely determined by the values of F found from FFTs and the positions of the peaks from plots of $S$ vs. $B^{-1}$. The low frequency $F^1$ and $F^2$ oscillations give peaks at low integers approaching to the origin of the plots. We also prepared the Landau level fan diagram for $F^4$ (shown in the lower inset in Fig. 7) for comparison with previous works. Having found the values of $n$, all data, apart from $F^1$ and $F^5$ that were excluded due to the small number of maxima available for indexing, were then fitted to straight lines with two free parameters, where, as shown in the figures, the resulting slopes match well with the quantum oscillation frequencies determined from FFT. We obtain the extrapolated phase shift of $F^4$ , $n(B^{-1} = 0)$ = -0.23 ± 0.02 that agrees perfectly with values given by X. Wang et al. [6] as well as M.N. Ali et al. [13] on the basis of SdH measurements, since quantum oscillations in the diffusion thermoelectric power are shifted by $\pm\pi/2$ (or $\delta n = 0.25$) in relation to SdH [25-27]. This, along with the fact that we do not observe any variation of the phase of the oscillations with the magnetic field, indicates that $S$ remains in the so called Mott regime [27]. This $\pi/2$ phase shift arises because the diffusion thermoelectric power depends on how the electron or hole density of states changes with energy. According to the Mott formula [28], $S$ is related to the logarithmic derivative of the electrical conductivity ($\sigma$) with respect to energy ($\varepsilon$) at the chemical potential ($\mu$) by:

$$S = -\frac{\pi^2}{3}\frac{k_B^2 T}{e}\left(\frac{\partial \ln\sigma(\varepsilon)}{\partial \varepsilon}\right)_\mu. \qquad (2)$$

Taking into consideration the sign of the charge carriers, we conclude that the shift is $-\pi/2$ for electrons and $+\pi/2$ for holes. Making the usual assumption that SdH oscillations arise because changes in $\sigma$ are proportional to oscillations in the density of states, then the total phase shift ($\phi$) in the Lifshitz-Kosevich formula describing the SdH effect: $\cos[2\pi(F/B + \phi)]$ [23] is believed to be $\pm 1/2$ and $\pm 5/8$ for a 2D and 3D parabolic band, respectively (+ for holes and − for electrons) [29], whereas 0 and $\pm 1/8$ are expected for 2D and 3D Dirac cones [29] because of the additional Berry phase $\pi$ that is accumulated by Dirac fermions along cyclotron orbits [30,31]). Since the phase shifts measured in the thermoelectric power are: -1/4 (-0.23 ± 0.02), 3/8 (0.38 ± 0.03), and 0 (0.04 ±0.04), (in units of $n$ i.e. $2\pi$) for $F^4$, $F^3$, $F^1$ respectively, then, if solutions indicating contradictory signs of phase shifts in $S$ and $\sigma$ are excluded, for $F^4$ one has: $\phi = -1/4 + 1/4 = 0$ which is expected for a 2D electron-like Dirac cone, and for $F^3$: $\phi = 3/8 - 1/4 = 1/8$, which is expected for a 3D hole-like Dirac cone. For $F^1$ we obtain $\phi = 0 \pm 1/4 = \pm 1/4$, which is rather unexpected. Therefore we conclude that $F^3$, with the phase shift being exactly 1/8, very small effective mass, and high mobility, that allows oscillations to be visible even at $T = 100$ K, is a textbook example of three dimensional Dirac fermion behaviour. We also indicate that $F^4$ is a 2D Dirac cone which is consistent with electronic structure calculations and

ARPES results [4]. On the other hand, the phase shift determined for $F^1$ does not match values suggested by present theoretical models and remains puzzling.

In summary we show that the thermoelectric power is a powerful tool for studying the complex quantum properties of the ZrSiS nodal-line semimetal. We distinguish five different oscillations in $S(B^{-1})$ with frequencies from $F^1 = 8.5$ T to $F^5 = 583$ T, whereas measurements of the de Haas van Alphen and Shubnikov de Haas effects were only able to detect two of them simultaneously. By applying the Lifshitz-Kosevich formula modified for the thermoelectric power for all five modes we determined the effective masses, which for $F^1$, $F^2$, $F^3$ and $F^4$ turned out to be one or two orders of magnitude smaller than the free electron mass. Determining the phase shifts allowed us to conclude that $F^3$ and $F^4$ oscillations originate from hole-like 3D and electron-like 2D Dirac cones, respectively. On the other hand, we do not find a good explanation for the zero shift measured for $F^1$. This low frequency oscillation appears to reach the quantum limit when the magnetic field higher that 10 T.

**Methods**

Single crystals of ZrSiS were grown by a chemical vapour transport method described elsewhere [11]. The thermoelectric power ($S$) was measured along the $a$ axis of a single crystal of dimensions 1.7 x 1.7 x 0.12 mm$^3$ with magnetic field ($B$) parallel to the $c$ axis. A sample was clamped between two phosphor bronze blocks, to which two Cernox thermometers and resistive heaters were attached. The signal was measured at a given temperature during a field sweep from -12.5 to +12.5 T and vice versa, in order to exclude a small component of the thermoelectric effect that was antisymmetric in $B$. For measurements in zero field, corrections were made for the small contribution to $S$ from the phosphor bronze measuring wires.

**Acknowledgements**

We are grateful to Kamran Behnia for useful suggestions.

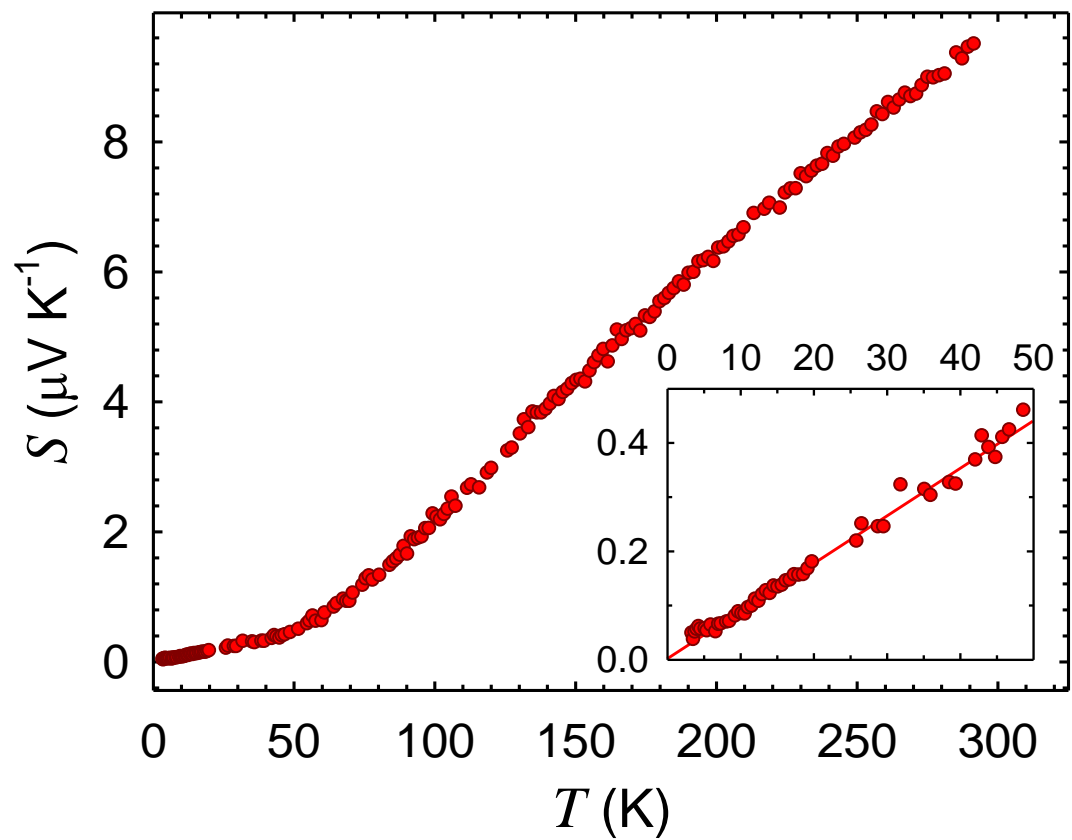

**Fig. 1.** The temperature dependence of the *a* axis thermoelectric power for ZrSiS.

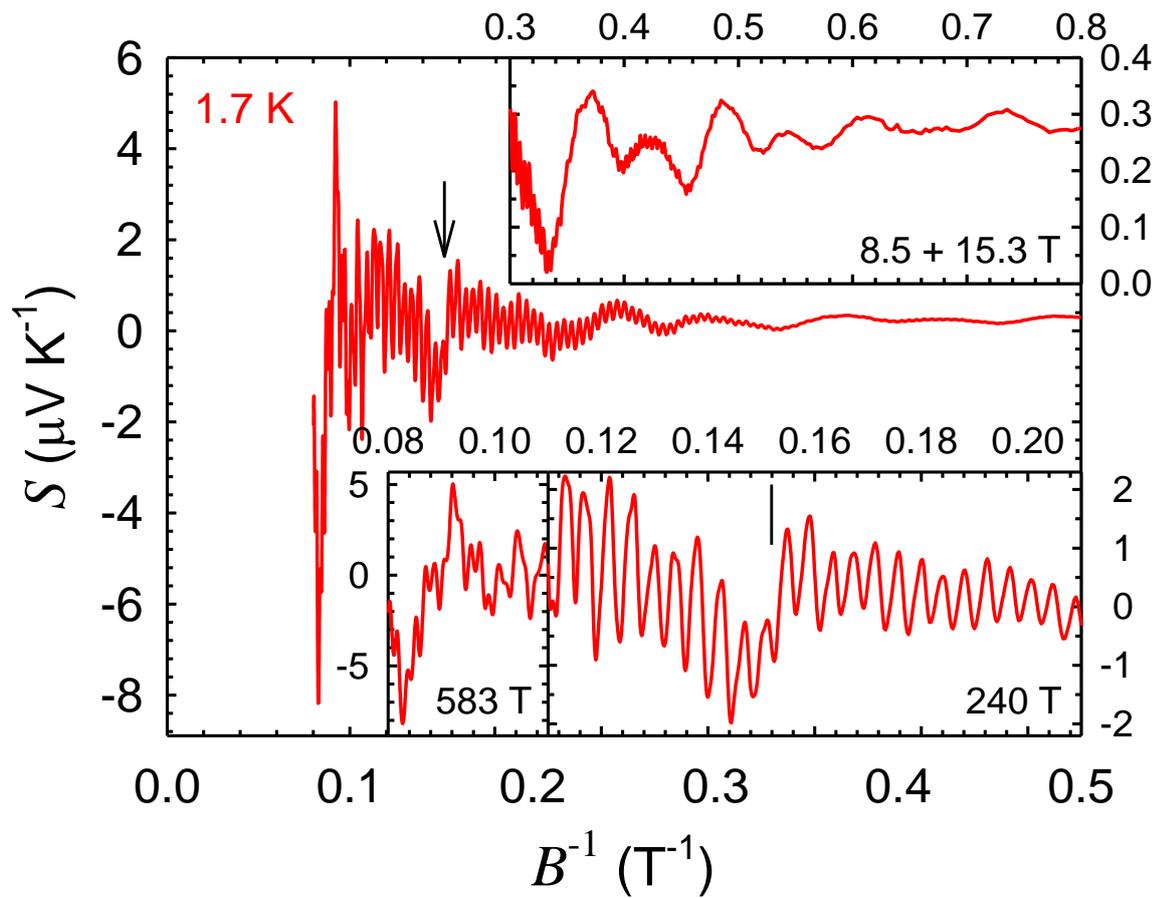

**Fig. 2.** The magnetic field dependence of the thermoelectric power in ZrSiS at T = 1.7 K. Insets show the same data on different scales. The corresponding oscillation frequencies (F in Tesla) are also shown. The arrows mark a jump at $B^{-1} = 0.15\ T^{-1}$ discussed in the text.

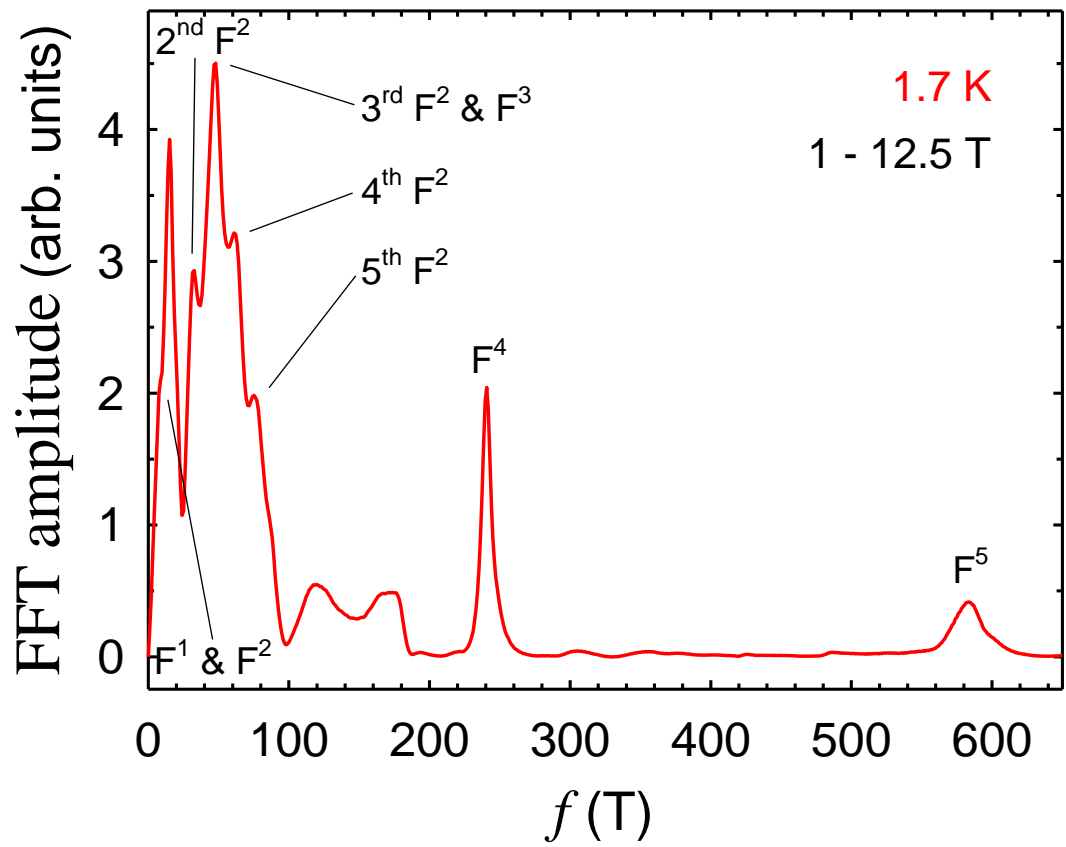

**Fig. 3.** The Fast Fourier Transform spectrum calculated for the thermoelectric power data at $T = 1.7$ K.

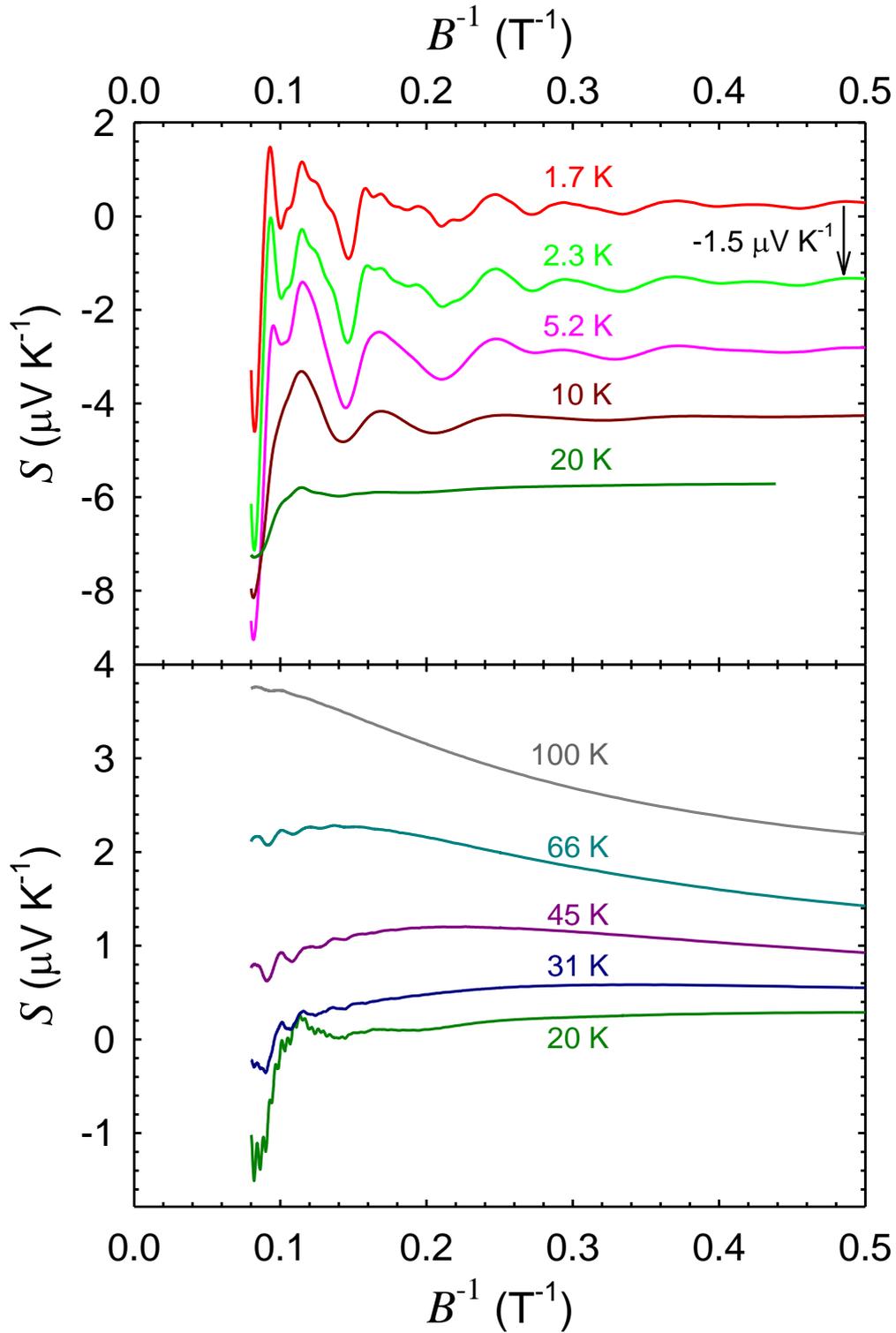

**Fig. 4.** The magnetic field dependences of the thermoelectric power in ZrSiS at selected temperatures. Upper panel shows low temperature ($T \leq 20$ K) data (shifted vertically by 1.5 µV/K for clarity) where high frequency oscillations were filtered out. Bottom panel show the raw high temperature data ($T \geq 20$ K).

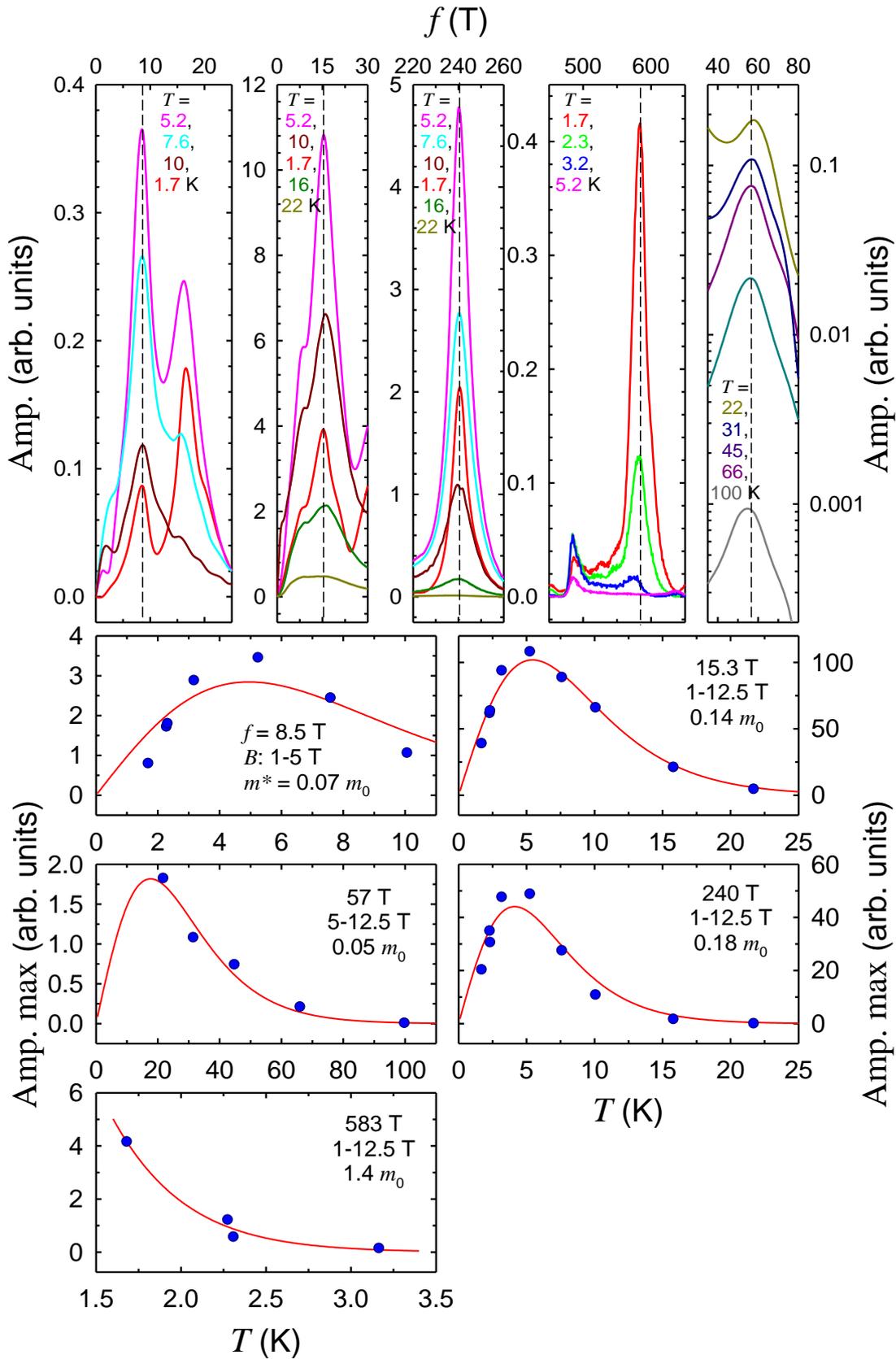

**Fig. 5.** The temperature dependences of the peaks height obtained from FFT. The solid red lines in the bottom panels are fits to Eqn. 1.

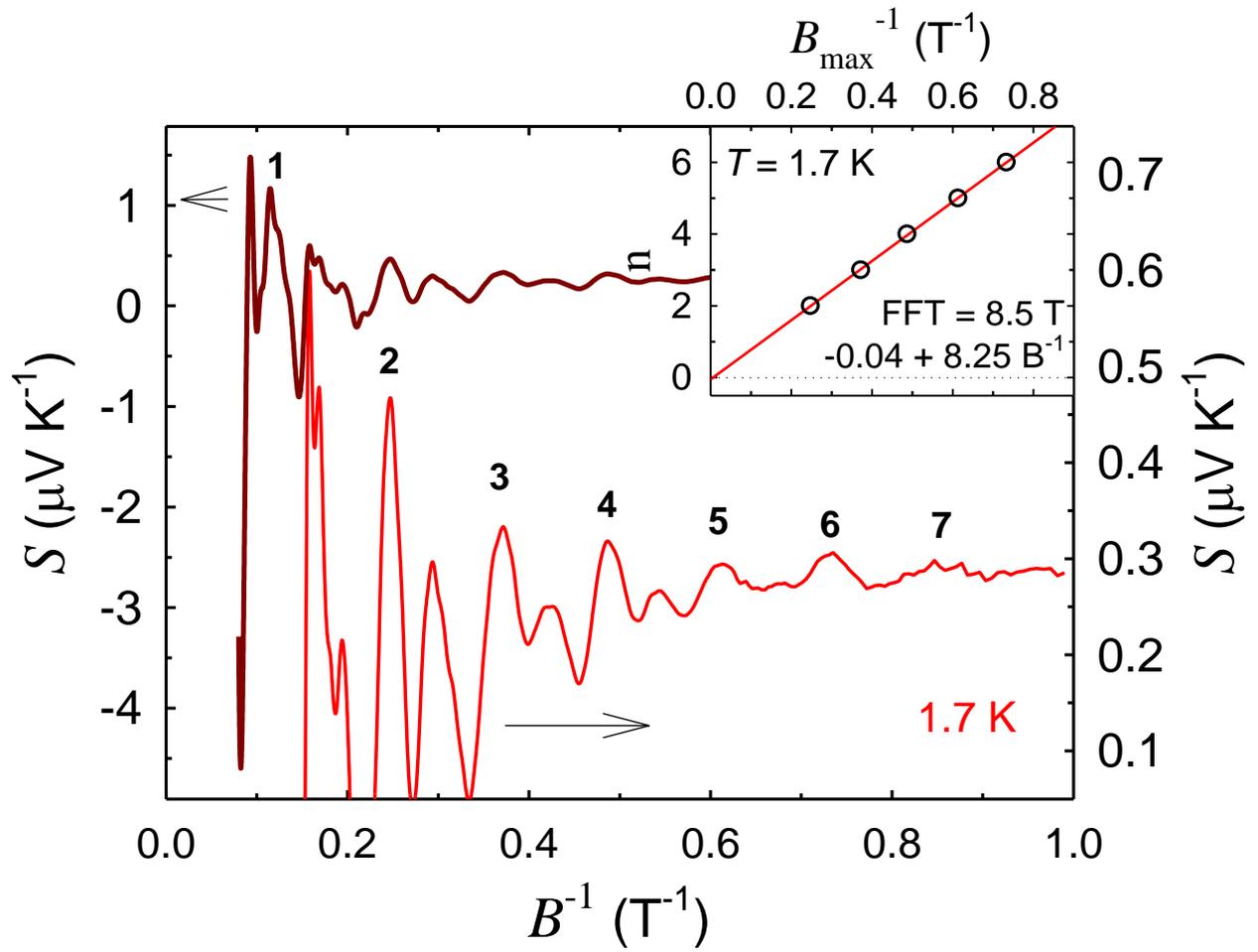

**Fig. 6.** The magnetic field dependence of the filtered (low frequency) thermoelectric power data in ZrSiS at 1.7 K. Inset shows the Landau level index plot for $F^1$ (8 T).

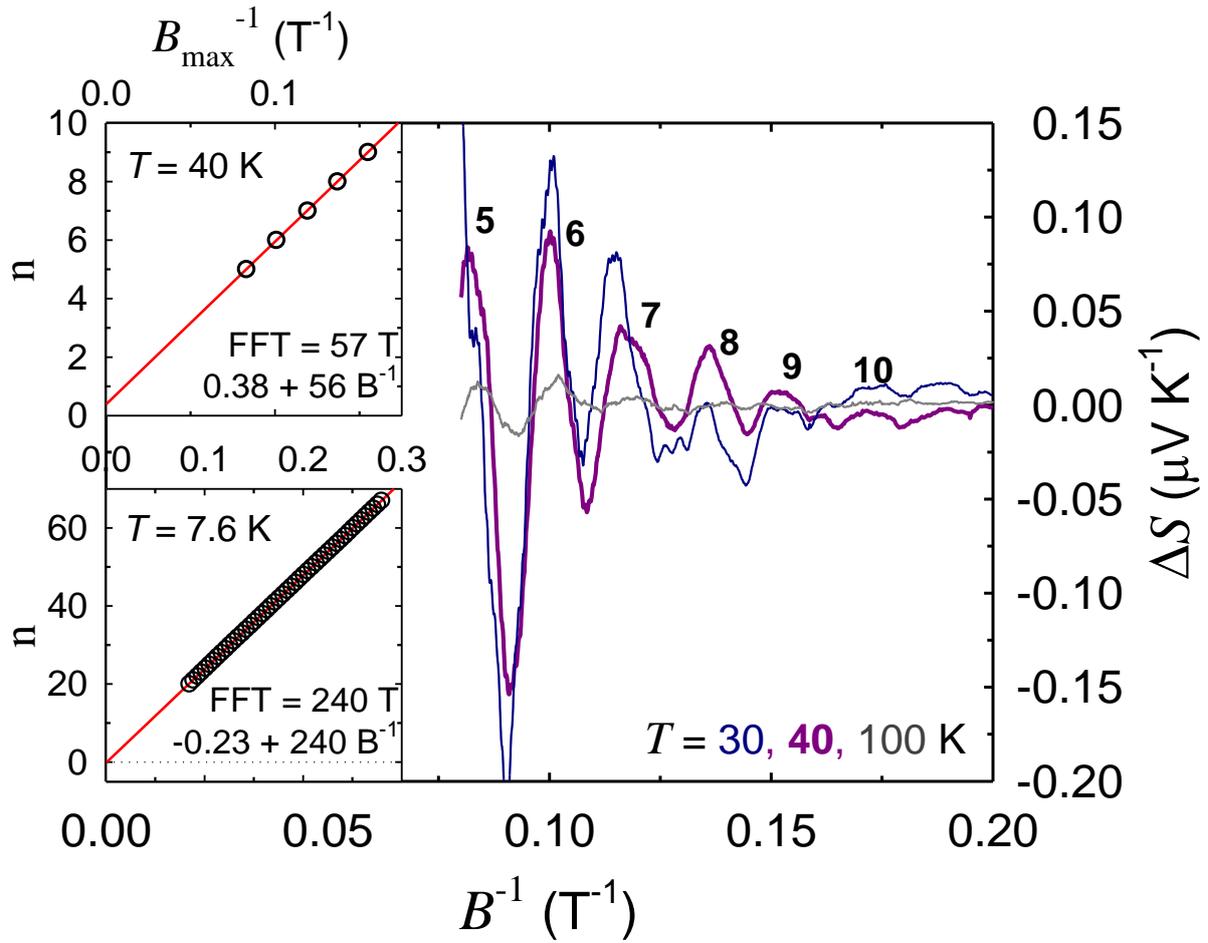

**Fig. 7.** The magnetic field dependence of the oscillatory part of the thermoelectric power in ZrSiS at $T$ = 30, 40 and 100 K. Insets show the Landau level index plot for $F^3$ (upper one, 57 T) and $F^4$ (bottom one, 240 T).